\documentclass[secnumarabic, graphics,floatfix, nofootinbib,tightenlines,nobibnotes, aps, prl, 12pt]{revtex4-1}
\usepackage{graphicx}
\usepackage[english]{babel}
\usepackage[utf8]{inputenc}
\usepackage{amsmath,amssymb}
\usepackage{amsfonts}
\usepackage{multirow}

\def\be{\begin{equation}}
\def\ee{\end{equation}}
\def\bea{\begin{eqnarray}}
\def\eea{\end{eqnarray}}

\begin{document}

\title{p-wave holographic superconductor in scalar hairy black holes}

\author{Dan Wen$^{1}$}
\author{Hongwei Yu$^{1}$\footnote{hwyu@hunnu.edu.cn}}
\author{Qiyuan Pan$^{1}$}
\author{Kai Lin$^{2,3}$}
\author{Wei-Liang Qian$^{3,4,5}$}

\affiliation{$^{1}$ Key Laboratory of Low Dimensional Quantum Structures and Quantum Control of Ministry of Education, Synergetic Innovation Center for Quantum Effects and Applications, and Department of Physics, Hunan Normal University, Changsha, Hunan, China}
\affiliation{$^{2}$ Hubei Subsurface Multi-scale Imaging Key Laboratory, Institute of Geophysics and Geoinformatics, China University of Geosciences, 430074, Wuhan, Hubei, China}
\affiliation{$^{3}$ Escola de Engenharia de Lorena, Universidade de S\~ao Paulo, 12602-810, Lorena, SP, Brazil}
\affiliation{$^{4}$ Faculdade de Engenharia de Guaratinguet\'a, Universidade Estadual Paulista, 12516-410, Guaratinguet\'a, SP, Brazil}
\affiliation{$^{5}$ School of Physical Science and Technology, Yangzhou University, 225002, Yangzhou, Jiangsu, P.R. China}

%\date{\today}
\date{March 07, 2019}

\begin{abstract}
We study the properties of the p-wave holographic superconductor for the scalar hairy black holes in the probe limit. 
The black hole solutions in question possess planar topology, which are derived from the Einstein gravity theory minimally coupled to a scalar field with a generic scalar potential. 
These solutions can be viewed as characterized by two independent parameters, namely, $\alpha$ and $k_0$, where AdS vacuum is manifestly restored when $\alpha\to \infty$.
Consequently, the p-wave holographic superconductor is investigated by employing the above static planar black hole spacetime as the background metric, where a Maxwell field is introduced to the model by  nonminimally  coupling it to a complex vector field.
The latter is shown to condensate and furnish the superconducting phase when the temperature is below a critical value.
By numerical calculations, we examine in detail how the scalar field in the background affects the properties of the superconductivity. 
It is found that the critical temperature depends crucially on the parameters $\alpha$ and $k_0$, which subsequently affects the condensation process. 
By employing the Kubo formula, the real, as well as imaginary parts of the conductivity, are calculated and presented as functions of frequency. 
The results are discussed regarding the poles of the Green function, and the typical values of the BCS theory.

\end{abstract}

\maketitle

%%%%%%%%%%%%%%%%%%%%%%
\section{I Introduction}
%%%%%%%%%%%%%%%%%%%%%%

The gauge/gravity duality is a conjectured relationship between quantum field theory and gravity.
Owing to its non-perturbative nature, the duality is regarded as a significant advance concerning both the string theory and quantum gravity.
In its original formulation, the concept of holographic principle was first proposed by 't Hooft~\cite{hooft1993dimensional} which states that all the physical phenomena of a volume of space can be described entirely by the degrees of freedom residing on the boundary of the region.
Then the idea is further promoted by Susskind via a string theory interpretation~\cite{susskind1995world}.
As a rigorous realization of the holographic principle, the anti-de Sitter/conformal field theory (AdS/CFT) correspondence was first proposed by Maldacena~\cite{maldacena1999large}, and consequently, many vital aspects were further developed.
The duality is mostly viewed as a powerful tool due to the fact that it relates the physics of a strongly interacting many-body system to that of classical gravity in higher dimensions with weak coupling.
In other words, it transforms difficult problems in the former arena into mathematically more tractable ones in the latter.
As a result, the duality provides a possibility to explore systems with strong coupling, not exclusively to gravitational applications.
For instance, it has been employed to investigate different aspects in nuclear~\cite{adscft-rhic-01} and condensed matter~\cite{adscft-condensed-review-02} physics.

Regarding condensed matter systems, one of the most promising topics that the theory has been applied to concerns nonconventional superconductors~\cite{adscft-condensed-review-02}.
By considering a charged AdS black hole coupled to a massive charged scalar field, Gubser~\cite{gubser2008breaking} shows that the gauge invariance is spontaneously broken, which leads to condensation of the charged scalar field.
The above model furnishes a simple gravitational system that admits black holes with charged ``scalar hair" which disappears at high temperature.
The dual system is subsequently interpreted as the so-called s-wave holographic superconductor model via the AdS/CFT correspondence~\cite{hartnoll2008building}.
The resultant model shows that gauge/gravity duality provides new insight into the physical mechanism of the high-temperature superconductors which cannot be understood in the framework of BCS theory.
Subsequently, the model is generalized to describe d-wave superconductors through the condensation of a charged massive spin-two field in the bulk~\cite{chen2010towards,benini2010gauge}.
Furthermore, models for p-wave superconductors are also developed by introducing a charged vector field~\cite{mackenzie2000p,cai2013holographic}.
The related topics have since been extensively investigated~\cite{horowitz2011introduction,herzog2009lectures,Lin:2016swr,Pan:2015lit}, including topics related to superfluids~\cite{Wu:2014bba,Lan:2017qxm}.

On the other hand, black hole has been demonstrated to play an essential role in relating various distinct subjects.
For the study of holographic superconductor, in particular, the concept of temperature in the dual system is realized by the introduction of a planar black hole into the bulk.
By investigating different background black hole metrics, one is allowed to access different physical systems associated with the dual counterparts on the gauge theory side, and particularly, to obtain substantial insight for realistic systems of physical interest~\cite{cai2013holographic}.
For instance, the inclusion of the higher curvature contributions, such as those in Gauss-Bonnet as well as its generalization Lovelock garvities, allows one to explore the properties of the dual system when the coupling is less strong~\cite{Pan:2011ah,Cai:2010cv,Gangopadhyay:2012np}.
The p-wave holographic superconductor was studied for a particular Lovelock gravity~\cite{wen2018maxwell} in comparison with that for the Gauss-Bonnet gravity~\cite{lu2016holographic}.
The superconducting phase dual to a solution in asymptotically AdS space~\cite{Nishioka:2009zj} is also interesting owing to the transition from the AdS black hole to the AdS soliton in the Poincar\'e coordinate while maintaining the boundary of the bulk~\cite{Surya:2001vj}.
Also, theories of modified gravity have also been explored such as those of Horava-Lifshitz theory~\cite{Wu:2014bba,Mansoori:2016zbp}.

Recently, several new classes of solutions for scalar hairy black holes in asymptotically AdS spacetimes have been investigated by Fan {\it et al.}~\cite{fan2015static,fan2016exact}.
These solutions are derived from Einstein gravity coupled to a scalar field, either minimally or non-minimally, with  a generic scalar potential.
They explicitly provide analytical examples for the formation of scalar hairy black holes, which further strengthen the ongoing research concerning the counterexamples of the no-hair theorems in various spacetimes.
It is found~\cite{fan2016exact} that a specific class of static hairy black hole solutions in asymptotically AdS spacetimes is controlled by a unique independent integration constant $\tilde{q}$, which does not appear in  the scalar potential.
The parameter can be associated with the ``charge" of the scalar field~\cite{fan2015static}, which in turn characterizes the properties of the black hole solutions such as temperature or entropy.
Moreover, the dynamic analytic solutions are explored by the authors by promoting the scalar charge to be dependent on the advanced or retarded times in the Eddington-Finkelstein like coordinates.
In respect of the AdS/CFT correspondence, these solutions broaden the possible physical scenarios that can be studied dual to the gravity side.
In particular, the constructed black hole metric involving a minimal coupling of the scalar field is compact in its physical configuration while elegant in mathematical form, and therefore provides an attractive candidate for the holographic superconductor.

The present study involves an attempt to construct a p-wave holographic superconductor by making use of the above hairy planar black hole solution as the background metric.
In order to realize a p-wave symmetry, on the top of the background metric in the bulk, the model introduces a complex vector field non-minimally coupled to a local U(1) gauge field, the Maxwell field.
The latter is dual to a strongly coupled system concerning a charged vector operator with a global U(1) symmetry on the boundary~\cite{cai2015introduction}.
In the probe limit, we numerically study the properties of the vacuum expectation value of the dual vector operator, and thus the superconductor phase transition.

The rest of the paper is organized as follows.
In Section II, we construct the p-wave holographic superconductor model in the scalar hairy black hole metric.
The equations of motion and the asymptotic solution at the boundary are derived.
In Section III, we present the numerical results by showing the condensate as a function of the temperature and its dependence on the parameters of the black hole.
Also, the roles of the mass of the vector field are investigated.
In Section IV, we investigate the real and imaginary parts of the conductivity as a function of the ratio of the frequency to the critical temperature.
Finally, section V is devoted to the discussions and concluding remarks.

%%2%%%%%%%%%%%%%%%%%%%%
\section{II The model} \label{Sect2}
%%%%%%%%%%%%%%%%%%%%%%%%%%%

Recently, Fan and Chen obtained analytic hairy planar black hole solutions in Einstein gravity, where a real scalar field $\Phi$ is minimally coupled to the metric with potential $V(\Phi)$~\cite{fan2016exact}.
For the specific form of the potential given by
\bea
V(\Phi) &=&  -\frac{1}{2}(\mathrm{cosh}\Phi)^{\frac{\tilde{\mu} k_0^2}{d-2}}\Big[g^2-\alpha(\mathrm{sinh}\Phi)^{\frac{d-1}{\tilde{\mu}}} {_2F_1}(\frac{d-1}{2\tilde{\mu}}, \frac{\tilde{\mu} k_0^2}{4(d-2)}, \frac{d+2\tilde{\mu}-1}{2\tilde{\mu}}, -(\mathrm{sinh}\Phi)^2)\Big] \nonumber \\
&&\times(2(d-2)(d-1)-\tilde{\mu}^2k_0^2(\mathrm{tanh}\Phi)^2)-\alpha(d-2)(d-1)(\mathrm{cosh}\Phi)^{\frac{\tilde{\mu} k_0^2}{2(d-2)}} (\mathrm{sinh}\Phi)^{\frac{d-1}{\tilde{\mu}}}  , \nonumber\\
\eea
with
\bea
\Phi = \frac{\tilde{\phi}}{k_0} ,
\eea
the static planar black hole metric reads
\bea
ds^2 &=& -r^2fdt^2+\frac{\sigma^2 dr^2}{r^2f} +r^{2}d\vec{x}^2_{d-2}  \, , \label{BH} \\
\tilde{\phi} &=& k_0 \mathrm{arcsinh} (\frac{\tilde{q}^{\tilde{\mu}}}{r^{\tilde{\mu}}})           \, , \label{scalar} \\
f &=& g^2-\frac{\alpha \tilde{q}^{d-1}}{r^{d-1}} {_2F_1}(\frac{d-1}{2\tilde{\mu}}, \frac{\tilde{\mu} k_0^2}{4(d-2)}, \frac{d+2\tilde{\mu}-1}{2\tilde{\mu}}, -\frac{\tilde{q}^{2\tilde{\mu}}}{r^{2\tilde{\mu}}})\, , \label{fexpr} \\
\sigma &=& (1+\frac{\tilde{q}^{2\tilde{\mu}}}{r^{2\tilde{\mu}}})^{-\frac{\tilde{\mu} k_0^2}{4(d-2)}} \label{sigma},
\eea
where $d\vec{x}^2_{d-2}$ represents the line element in $(d-2)$ dimensional space with planar topology.
$f$, $\sigma$ and the scalar $\tilde{\phi}$ are functions of $r$, and $_2F_1$ is the hypergeometric function.
Here $g^2$ is a numerical constant, while $\alpha$, $k_0$ and $\tilde{q}$ are model parameters.
In particular, $\tilde{q}$ is associated with the scalar field, which can be loosely interpreted as a non-conserved ``charge".
The scalar mass $m_0$ satisfies $m_0^2=-\tilde{\mu}(d-\tilde{\mu}-1)g^2$.
In the following calculations, we focus on the case where $\tilde{\mu}=\frac{d-1}{2}$ and $g=1$ in $d=5$ spacetime.

By using the condition of $f(r_h)=0$ with $r_h=1$, we obtained the relation between the model parameters $\tilde{q}$, $k_0$, and $\alpha$
\begin{eqnarray}
\tilde{q}=[-1+(1 +\frac{1 -\frac{k_0^2}{6}}{\alpha})^{\frac{-6}{6+k_0^2}}]^{\frac{1}{4}}  \, . \label{q4}
\end{eqnarray}
Therefore, the background metric is determined by two free parameters, which we choose to be $\alpha$ and $k_0$.
When the parameter $\alpha$ increases as $\alpha \to \infty$, one finds $\tilde{q} \to 0$ and $\alpha \tilde{q}^4 \to 1$, and subsequently due to Eqs.(\ref{fexpr}) and (\ref{sigma}), $f \to 1-\frac{1}{r^4}$ and $\sigma \to 1$.
This implies that the black hole solution restores to that of a Schwarzschild-AdS one in this limit.

The temperature and mass of the black hole are determined by
\bea
T = \frac{(d-1)\alpha \tilde{q}^{d-1}}{4\pi r_h^{d-2}}    \, , \label{Temp} \\
M = \frac{(d-2)\alpha \tilde{q}^{d-1}}{16\pi}  \, , \label{Mass}
\eea
where $r_h$ is the black hole horizon.

To build the holographic p-wave model of a superconductor using the above background metric, we consider the following matter action which includes a Maxwell field and a complex vector field~\cite{cai2015introduction}:
\begin{eqnarray}
\mathcal{I_{MCV}} &=& \frac{1}{16\pi G} \int d^dx \bigg\{\sqrt{-g}\Big[ -\frac{1}{4}F_{\mu\nu}F^{\mu\nu}-\frac{1}{2}(D_\mu\rho_\nu-D_\nu\rho_\mu)^\dagger(D^\mu\rho^\nu-D^\nu\rho^\mu) \nonumber \\
&& -m^2\rho^\dagger_\mu\rho^\mu+iq\gamma\rho_\mu\rho^\dagger_\nu
F^{\mu\nu}\Big] \bigg\} \, , \label{act}
\end{eqnarray}
where $F_{\mu\nu}=\nabla_\mu A_\nu-\nabla_\nu A_\mu$ is the Maxwell field strength tensor and the covariant derivative is given by $D_\mu=\nabla_\mu- i q A_\mu$.
The constants $q$ and $m$ represent the charge and the mass of the vector field $\rho_\mu$ respectively.
The last term in the action corresponds to the nonminimal coupling between the vector field $\rho_\mu$ and the gauge field $A_\mu$, where $\gamma$ is the magnetic moment of the vector field.
The present study will adopt the probe limit and ignore the backreaction to the metric.
We will also neglect the effect of the magnetic field on the superconductor transition.

From the action (\ref{act}), one finds the following equations of the motion for the matter fields
\begin{eqnarray}
D^\nu(D_\nu\rho_\mu-D_\mu\rho_\nu)-m^2\rho_\mu = 0\, , \label{eq1} \\
\nabla^\nu
F_{\nu\mu}-iq[\rho^\nu(D_\nu\rho_\mu-D_\mu\rho_\nu)^\dagger-\rho^{\nu\dagger}(D_\nu\rho_\mu-D_\mu\rho_\nu)]
= 0 \, .\label{eq2}
\end{eqnarray}
Here, we adopt the following ansatz for the solution of the vector field $\rho_\mu$
\begin{eqnarray}
\rho_\mu dx^\mu = \rho_x(r) dx \, , \
\end{eqnarray}
and \begin{eqnarray}
A_\mu dx^\mu =\phi(r)dt \, , \
\end{eqnarray}
for the gauge field $A_\mu$.
It is straightforward to show that the above equations of motion can be simplified to,
\begin{eqnarray}
\rho_x''+ \left[\frac{f'}{f}-\frac{\sigma'}{\sigma}+\frac{(d-2)}{r}\right]\rho_x'+\left[\frac{\phi^2}{r^4f^2}-\frac{m^2}{r^2f}\right]\sigma^2\rho_x = 0 \, , \label{eom1} \\
\phi''+ (\frac{d-2}{r}-\frac{\sigma'}{\sigma})\phi'-\frac{2\rho_x^2}{r^4f}\sigma^2 \phi = 0 \, . \label{eom2}
\end{eqnarray}

The boundary conditions at the horizon $r = r_h$ of the AdS$_d$ bulk are determined by requiring $\rho_\mu$ being regular, while the gauge field $A_\mu$ satisfies $\phi(r_h) =0$.
One can show that the asymptotic forms of the matter field and the gauge field are given by
\begin{eqnarray}
\rho_x(r) = \frac{\rho_{x-}}{r^{\Delta_-}}+\frac{\rho_{x+}}{r^{\Delta_+}}+\cdots \, ,\label{asympsi} \\
\phi(r) = \mu-\frac{\rho}{r^{(d-3)}}+\cdots\, ,\label{asymphi}
\end{eqnarray}
at the conformal boundary $r\rightarrow\infty$.
Here $\Delta_\pm = \frac{1}{2}[(d-3)\pm \sqrt{(d-3)^2+4m^2}]$, with the Breitenlohner-Freedman (BF) bound $m^2\geq -\frac{(d-3)^2}{4}$.
This indicates that when $m^2= -\frac{(d-3)^2}{4}=m_{BF}^2$, $\Delta_+=\Delta_-=\Delta_{BF}=\frac{d-3}{2}$.
The coefficients $\mu, \rho$ and $\rho_{x-}, \rho_{x+}$ are interpreted respectively, as the chemical potential, charge density, the source and the $x$ component of the vacuum expectation value of the dual vector operator $<\mathcal{O}>$ in question.
Besides, we impose the condition $\rho_{x-} = 0$ to guarantee that the vector condensation will arise spontaneously in the boundary theory.

Since the equations of motion Eqs.(\ref{eom1}) and (\ref{eom2}) satisfy the following scaling law:
\begin{eqnarray}
r \to \lambda r ,\       (t,x,y,\cdots) \to \frac{1}{\lambda}(t,x,y,\cdots) \, ,\ \\
  \rho \to \lambda^{d-2} \rho ,\    (T,\mu) \to \lambda (T,\mu),\        \rho_{x+} \to \lambda^{1+\Delta_+}\rho_{x+} \, ,\
\end{eqnarray}
with a positive constant $\lambda$, we make use of this fact to always choose $r_h = 1$ throughout the calculations and present the results in terms of dimensionaless quantities.
Now we are in a position to numerically solve the above equations of motion, Eqs.(\ref{eom1}) and (\ref{eom2}), by using the shooting method~\cite{hartnoll2008building,hartnoll2008holographic}.
As mentioned above, we use the scaling symmetries to set $r_h = 1$, and define $q=1$ without loss of generality.
The chosen boundary conditions imply that the solution of the above equations can be expressed in terms of two independent parameters determined at $r=r_h$, namely, $\phi'(r_h)$ and $\rho_x(r_h)$.
One of them is subsequently made use of to guarantee that the source term is vanishing on the horizon, $\rho_{x-} = 0$.
As a result, the other physical quantities, such as $\mu,~\rho,~\rho_{x+}$, can be obtained by reading off the corresponding coefficients of Eqs.(\ref{asympsi}) and (\ref{asymphi}) according to the AdS/CFT dictionary.
In order to evaluate the temperature dependence of the condensate, the scaling symmetry as mentioned above is made use of, and the results are expressed in terms of scale-invariant quantities, such as $T_c/\rho^{\frac{1}{d-2}}$ and $<\mathcal{O}>^{\frac{1}{1+\Delta_+}}/T_c$.
In this sense, the results acquired by varying the charge density $\rho$ for a given $r_h$ can be equivalently interpreted as those obtained by varying temperature $T$ for a given charge density $\rho$.

Also, one may investigate the properties regarding thermodynamical stability of the superconducting phase by evaluating the Gibbs free energy.
This is because the Gibbs free energy is minimized for a macroscopic stable thermodynamic system for given temperature and pressure, whereas the system can freely exchange heat and work with a large reservoir.
According to the AdS/CFT correspondence, the free energy can be evaluated in terms of the Euclidean on-shell action.
In particular, following Refs.\cite{adscft-superfluid-02,adscft-superfluid-04,adscft-superfluid-06}, one finds $\Omega=-\mathit{T}\mathcal{S}_{os}$, where $\mathcal{S}_{os}$ is the Euclidean on-shell action, namely,
\begin{eqnarray}
\mathcal{S}_{os}&=&\int dtdxdydzdr \sqrt{-g}\left[-\frac{1}{2}\nabla_{\mu}(A_\nu F^{\mu\nu}) -\nabla_{\mu}(\rho^\dagger_\nu \rho^{\mu\nu}) +\frac{1}{2}A_\nu \nabla_{\mu}F^{\mu\nu}\right] \nonumber \\
&=& \frac{V_3}{T}\left(-\frac{1}{2}\sqrt{-h}n_r A_\nu F^{r\nu}|_{r \to \infty} -\sqrt{-h}n_r \rho^\dagger_\nu \rho^{r\nu}|_{r \to \infty} + \frac{1}{2}\int^{\infty}_{r_h}dr\sqrt{-g} A_\nu \nabla_{\mu}F^{\mu\nu}\right) \nonumber \\
&=& \frac{V_3}{T}\left(\mu\rho -\int^{\infty}_{r_h} \frac{\sigma\rho^2_x\phi^2}{rf} dr\right).
\end{eqnarray}
Here $\int dtdxdydz=V_3/\mathit{T}$, and $\rho_{\mu\nu}=D_\mu \rho_\nu -D_\nu \rho_\mu$.
A factor $16\pi G$ is ignored for convenience.
The resultant free energy for the superconducting phase reads
\begin{eqnarray}
\frac{\Omega_S}{V_3}=-\frac{\mathit{T}\mathcal{S}_{os}}{V_3}=-\mu\rho +\int^{\infty}_{r_h} \frac{\sigma\rho^2_x\phi^2}{rf}dr .\label{gPsuper}
\end{eqnarray}
For the normal phase, one has $\rho_x=0$ and $A_t(r_h)=0$, and therefore
\begin{eqnarray}
\frac{\Omega_N}{V_3}=-\mu^2. \label{gPnormal}
\end{eqnarray}

%%%%%%%%%%%%%%%%%%%%%%%%%%%
\section{III Condensation of the vector operator} \label{Sect3}
%%%%%%%%%%%%%%%%%%%%%%%%%%%

By varying the two free parameters, $\alpha$ and $k_0$, of the metric, we study the effect of different black hole configuration on the condensation of the vector operator in $d=5$ dimensional spacetime.
Alternatively, one may express the results regarding a pair of variables regarding more straightforward physical quantities, namely, the scalar ``charge" $\tilde{q}$ and black holes mass $M$.
The latter can be evaluated by using Eqs.(\ref{fexpr}) and (\ref{Mass}).
To be specific, $\tilde{q}$ is determined by Eq.(\ref{q4}).
In Table \ref{tb1}, we show the calculated critical temperature by varying the values of $k_0$ and $\alpha$.
Considering that the mass of the vector field $\rho_{\mu}$ also affects condensation~\cite{wen2018maxwell}, the numerical results are also presented for different values of $m$.

\begin{table}[htb]
\begin{center}
  \caption{The calculated critical temperatures for different values of vector field mass $m$ as well as for parameters $\alpha, k_0$.}
  \begin{tabular}{c|c|c|c}
  \hline
            &   $d=5,m^2=-3/4$ &  $d=5,m^2=0$&  $d=5,m^2=5/4$\\
  \hline
   $k_0=\sqrt{3}, \alpha=7/5$&          0.248&        0.223&     0.205\\
  \hline
   $k_0=\sqrt{3}, \alpha=5$  &          0.231&        0.207&     0.190\\
  \hline
   $k_0=\sqrt{3}, \alpha=20$ &          0.226&        0.202&     0.185\\
  \hline
   $k_0=\sqrt{3}, \alpha=100$&          0.225&        0.201&     0.184\\
  \hline
   $k_0=\sqrt{3}, \alpha=400$ &         0.224&       0.201&      0.184 \\
  \hline
   $k_0=\sqrt{3}, \alpha=100000$ &      0.224&       0.201&      0.184\\
  \hline
   $k_0=2\sqrt{3}, \alpha=27/10$&       0.302&       0.272&      0.250\\
  \hline
   $k_0=2\sqrt{3}, \alpha=5$ &          0.258&       0.232&      0.213\\
  \hline
   $k_0=2\sqrt{3}, \alpha=20$ &         0.232&        0.207&     0.190\\
  \hline
   $k_0=2\sqrt{3}, \alpha=100$ &        0.226&        0.202&      0.185\\
  \hline
   $k_0=2\sqrt{3}, \alpha=400$ &        0.225&        0.201&      0.184\\
  \hline
   $k_0=2\sqrt{3}, \alpha=1000$ &       0.224&        0.201&      0.184\\
  \hline
   $k_0=2\sqrt{3}, \alpha=100000$&      0.224&        0.201&      0.184\\
  \hline
  \end{tabular}\label{tb1}
\end{center}
\end{table}

The critical temperature $T_c$ is inversely proportional to $\rho^{1/3}$ for $d=5$ dimensional spacetime.
First, $T_c/\rho^{1/3}$ increases as the mass of the vector field $m$ decreases.
Therefore, it becomes more difficult for the vector field to condense while its mass increases.
On the other hand, for a fixed value of $k_0$, the ratio $T_c/\rho^{1/3}$ decreases as $\alpha$ increases.
For a given value of $\alpha$, the ratio $T_c/\rho^{1/3}$ increases with increasing $k_0$.
In addition, we find that, independent of the value of $k_0$, the critical temperature approaches a constant as $\alpha$ increases.

\begin{figure}[ht]
\begin{center}
\begin{minipage}[c]{0.5\textwidth}
\centering\includegraphics[width=0.9\textwidth]{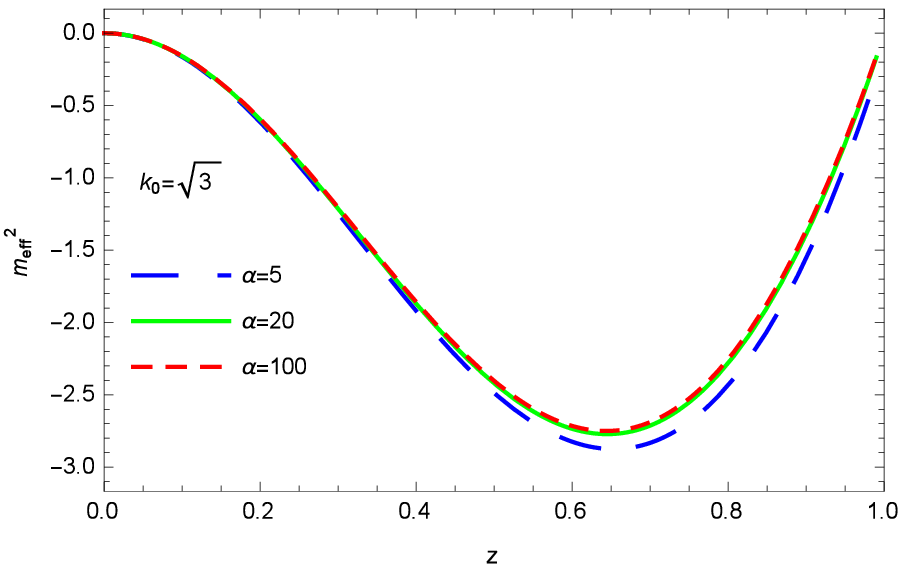}
\end{minipage}%
\begin{minipage}[c]{0.5\textwidth}
\centering\includegraphics[width=0.9\textwidth]{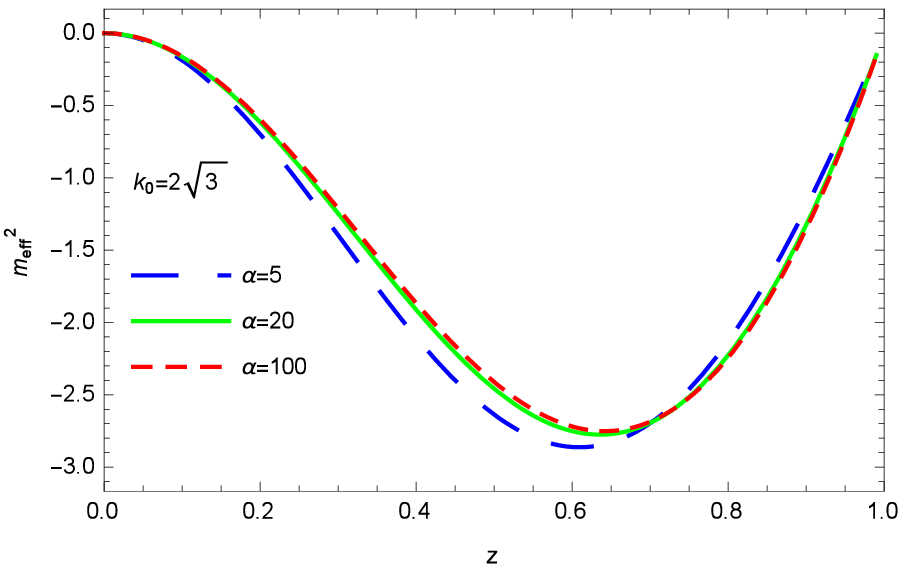}
\end{minipage}
\renewcommand{\figurename}{Fig.}
\caption{The calculated effective mass as a function of $z=\frac1r$, the mass of vector field is $m^2=0$.}
\label{effectivemasspotential}
\end{center}
\end{figure}

The effective mass~\cite{gubser2008breaking} of the vector field can be extracted from Eq.(\ref{act}), and one finds
\begin{eqnarray}
m^2_{eff} &=& m^2-\frac{\phi^2}{r^2f}   \, .\label{meff}
\end{eqnarray}
The resulting effective mass as a function of the radial coordinate is presented in Fig.\ref{effectivemasspotential}.
These plots provide information on how the condensation is affected by the background metric through the effective mass of the vector field.
From the resultant parameter dependence of the effective mass, information on the critcal temperature may be inferred.
In fact, for a given $k_0$, the calculated $m^2_{eff}$ decreases with decreasing $\alpha$, which results in higher critical temperature.
Besides, the effective mass does not depend on $k_0$ for large enough $\alpha$, and subsequently, the critical temperature is not affected by the model parameters in this limit.

The above findings can be understood as follows.
According to Eq.(\ref{meff}), the effective mass is a function of $f$ and $\phi$.
As discussed above, in the limit of $\alpha\to \infty$, the background metric reduces to the corresponding Schwarzschild-AdS one.
Furthermore, $\phi$ is a function of $r$ only in the case of the Schwarzschild-AdS limit.
As a result, effective mass is a function of $r$ only for a large $\alpha$.
Thus, the critical temperature becomes insensitive to $k_0$ when $\alpha$ assumes a large value.

\begin{figure}[ht]
\begin{center}
\begin{minipage}[c]{0.5\textwidth}
\centering\includegraphics[width=0.9\textwidth]{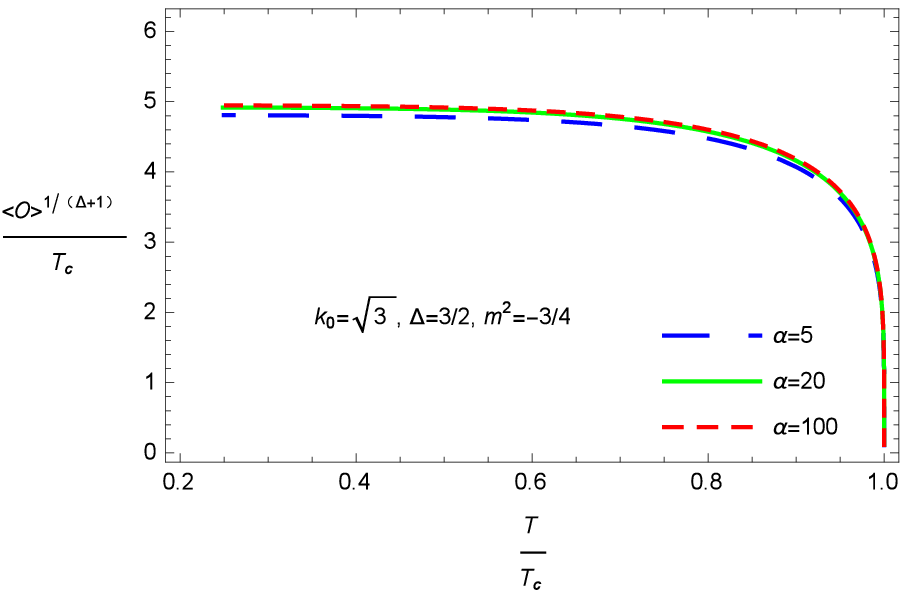}
\end{minipage}%
\begin{minipage}[c]{0.5\textwidth}
\centering\includegraphics[width=0.9\textwidth]{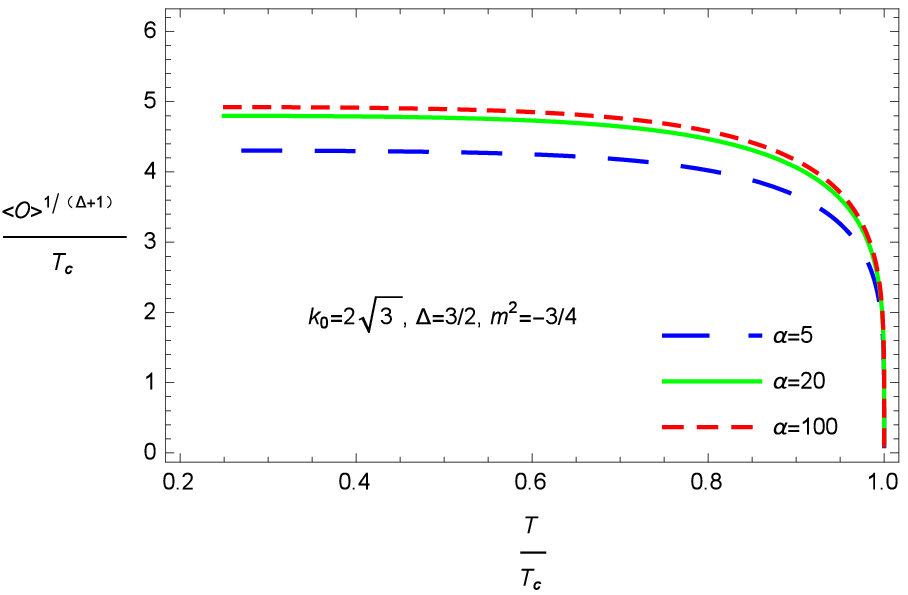}
\end{minipage}
\\
\begin{minipage}[c]{0.5\textwidth}
\centering\includegraphics[width=0.9\textwidth]{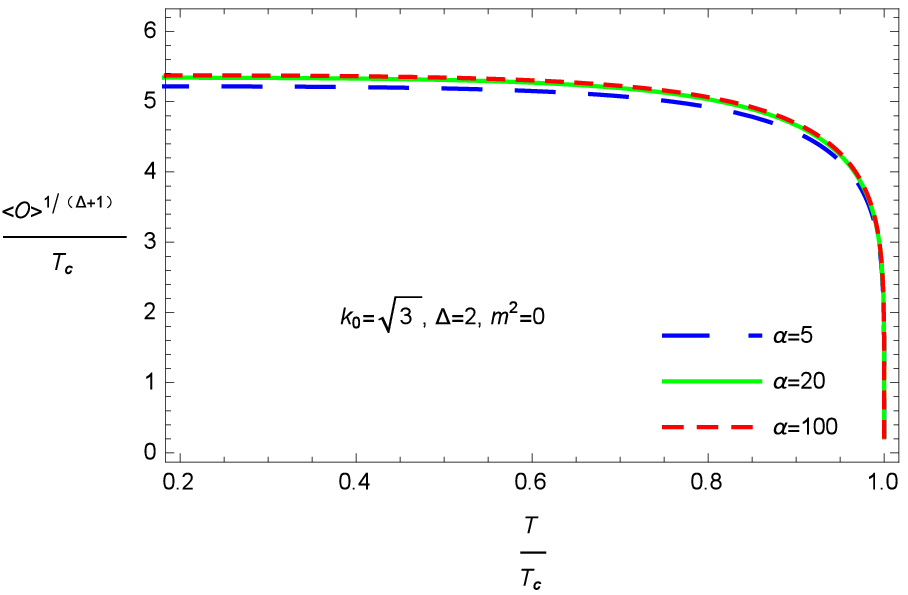}
\end{minipage}%
\begin{minipage}[c]{0.5\textwidth}
\centering\includegraphics[width=0.9\textwidth]{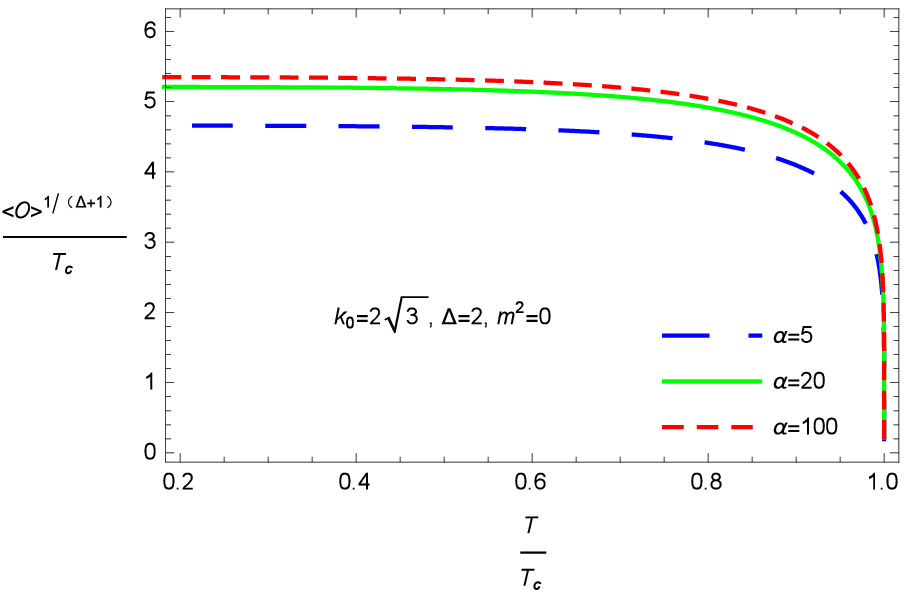}
\end{minipage}
\\
\begin{minipage}[c]{0.5\textwidth}
\centering\includegraphics[width=0.9\textwidth]{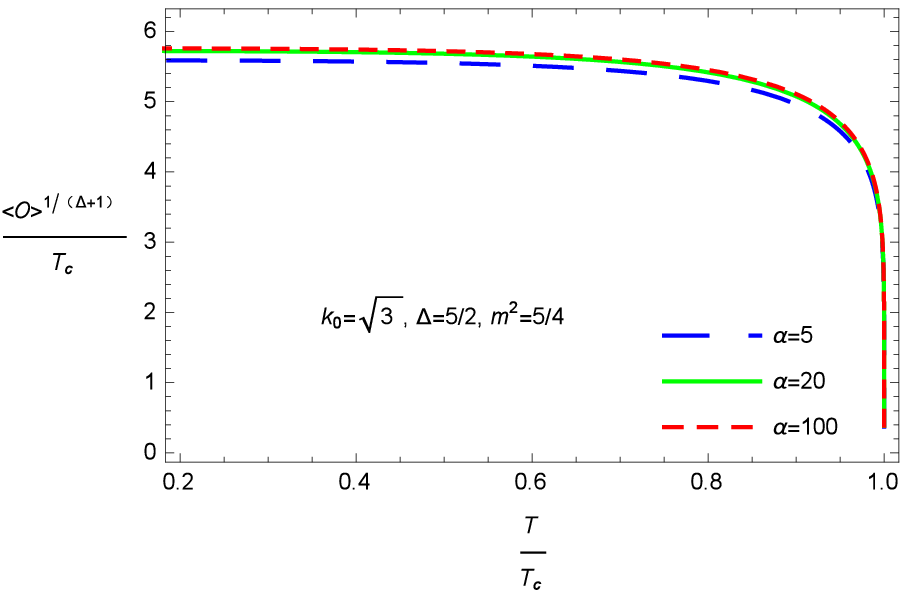}
\end{minipage}%
\begin{minipage}[c]{0.5\textwidth}
\centering\includegraphics[width=0.9\textwidth]{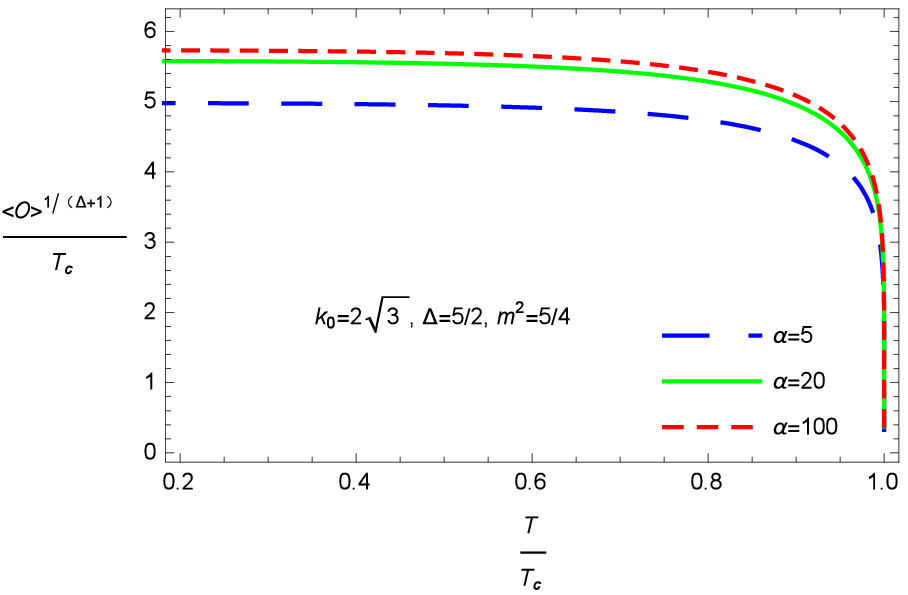}
\end{minipage}
\renewcommand{\figurename}{Fig.}
\caption{The calculated condensate as a function of temperature for different values of $\alpha$, $k_0$, $\Delta$ and $m$.}
\label{condensate1}
\end{center}
\end{figure}

In Fig.\ref{condensate1}, we show the calculated condensate as a function of the temperature for different values of $\alpha$, $k_0$ and $m$ in five-dimensional spacetime.
We first concentrate on the effects of the parameters of the background black hole.
The figure illustrates that the vector condensation becomes more significant as $\alpha$ increases for given $k_0$ and $m$.
This is consistent with the results shown in Tab.\ref{tb1}, where the critical temperature is found to decrease with increasing $\alpha$, as the condensation becomes more difficult for larger $\alpha$.
On the other hand, the condensation does not change significantly with $\alpha$, when $\alpha$ becomes larger than $20$, and it approaches a constant value as $\alpha\to \infty$.
For instance, for $k_0=\sqrt{3}$, clearly, the condensation curve for $\alpha=100$ almost coincides with that for $\alpha=20$, while the corresponding critical temperatures are found to be close to each other, as shown in Tab.\ref{tb1}.
For $\alpha=5$, the condensation for $k_0=\sqrt{3}$ is larger than that for $k_0=2\sqrt{3}$, and the critical temperature for $k_0=2\sqrt{3}$ is also more significant than that for $k_0=\sqrt{3}$.
As for $\alpha=100$, the difference for the condensations between $k_0=\sqrt{3}$ and $k_0=2\sqrt{3}$ vanishes, and the corresponding critical temperatures are consistent.
Thus one concludes that the transition to the superconductor phase becomes easier for a smaller $\alpha$ and a larger $k_0$.
On account of the fact that for large $\alpha$ the model restores to the Schwarzschild-AdS case, we conclude that the presence of the minimally coupled scalar field makes the superconductor transition easier than that of a pure Schwarzschild-AdS black hole.
Besides, we also investigate the effect of the vector field mass $m$ by repeating the calculations for different masses of the vector field $m^2=-3/4, 0, 5/4$ (with $\Delta_+=\Delta=3/2, 2, 5/2$ respectively).
One observes that as $m^2$ increases (and therefore $\Delta$ increases), the vector condensation becomes more significant with given $\alpha$ and $k_0$.

\begin{figure}[ht]
\begin{center}
\begin{minipage}[c]{0.5\textwidth}
\centering\includegraphics[width=1\textwidth]{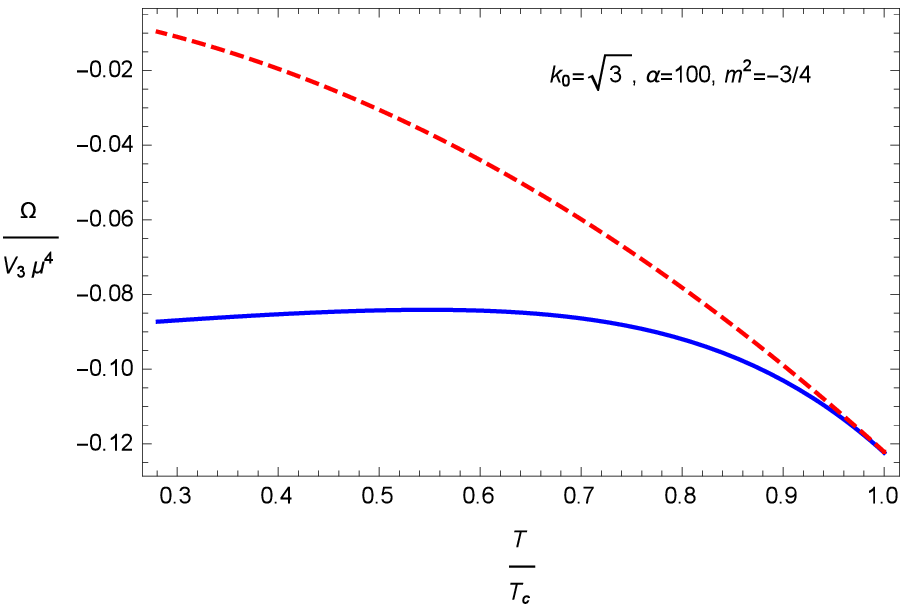}
\end{minipage}%
\begin{minipage}[c]{0.5\textwidth}
\centering\includegraphics[width=1\textwidth]{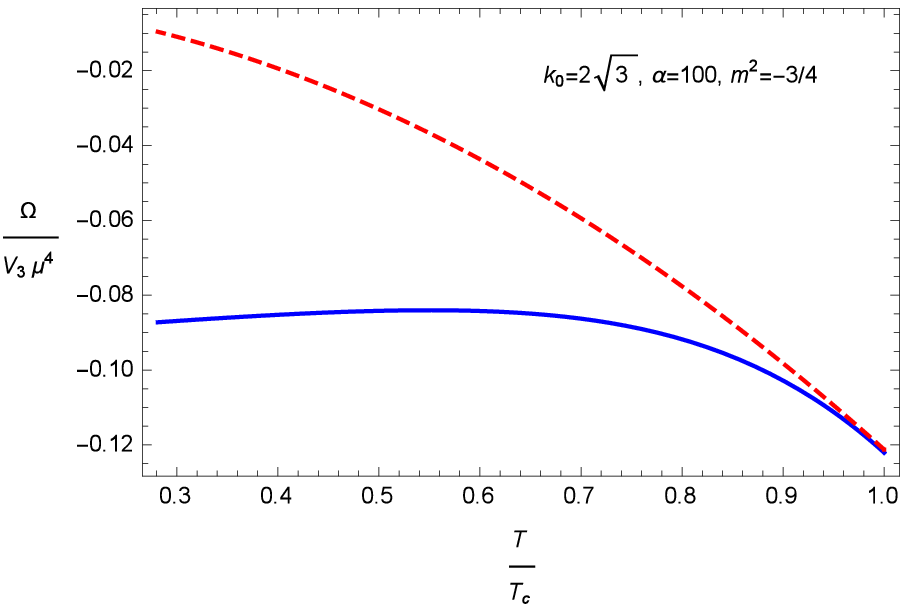}
\end{minipage}
\renewcommand{\figurename}{Fig.}
\caption{The calculated Gibbs free energy as a function of temperature for the superconducting phase (shown in blue solid curves) and normal phase (shown in red dashed curves).
The calculation are carred out for $k_0=\sqrt{3}, 2\sqrt{3}$ with $\alpha=100, m^2=-3/4$.
}
\label{freeEnergy}
\end{center}
\end{figure}

Besides, we evaluate the Gibbs free energies of the system for both the superconducting and normal phases by using Eqs.(\ref{gPsuper}) and (\ref{gPnormal}). 
Since other choices will not qualitatively change our results, we make use of the specific choices of parameters with $k_0=\sqrt{3}, 2\sqrt{3}$ with $\alpha=100$ and $m^2=-3/4$. 
The numerical results are presented in Fig.\ref{freeEnergy}.
It is found that the free energies, as well as their first order derivatives, numerically coincide at the critical temperature, as expected. 
Moreover, it is clearly shown that the free energy of the superconducting phase is smaller than that of the normal phase for $T< T_c$. 
As a result, one concludes that the formed superconducting phase is found to be thermodynamically favored.

%%%%%%%%%%%%%%%%%%%%%%%%%%%
\section{IV Conductivity} \label{Sect4}
%%%%%%%%%%%%%%%%%%%%%%%%%%%

In this Section, we numerically investigate the conductivity~\cite{gubser1998gauge,hartnoll2008building,gubser2008gravity,horowitz2011introduction,cai2013holographic,cai2015introduction,Chaturvedi:2015hra} by considering the perturbed Maxwell field $\delta A_y=A_y(r)e^{-i\omega t}dy$ in the case of the five-dimensional AdS black hole background.
The Maxwell equation reads
\begin{eqnarray}
A_y''+(\frac{f'}{f}-\frac{\sigma'}{\sigma}+\frac{3}{r})A_y'+(\frac{\omega^2}{r^4f^2}-\frac{2\rho_x^2}{r^4f})\sigma^2 A_y=0 , \label{eqAy}
\end{eqnarray}
with the ingoing wave boundary conditions at the horizon $(r = r_h =1)$
\begin{eqnarray}
A_y(r)\sim (r-1)^{-i\omega P} ,
\end{eqnarray}
where in the case of $k_0=\sqrt{3}$, we consider $P=\frac{5}{21}$ for $\alpha=5$, $P=\frac{20}{81}$ for $\alpha=20$, and $P=\frac{100}{401}$ for $\alpha=100$;
while for $k_0=2\sqrt{3}$ we consider $P=\frac{1}{5}$ for $\alpha=5$, $P=\frac{19}{80}$ for $\alpha=20$, and $P=\frac{99}{400}$ for $\alpha=100$.
The asymptotic behavior at infinity $r\rightarrow\infty$ is given by
\begin{eqnarray}
A_y(r)=A^{(0)}+\frac{A^{(2)}}{r^2}+\frac{A^{(0)}\omega^2}{2}\frac{\log r}{r^2}+\cdots .
\end{eqnarray}
In this case, it is well-known that the conductivity is related to the Green function via Kubo formula, which reads
\begin{eqnarray}
\sigma(\omega)=\frac{G^R}{i\omega}=-\frac{2i A^{(2)}}{\omega A^{(0)}}+\frac{i\omega}{2} ,
\end{eqnarray}
where the Green function satisfies
\begin{eqnarray}
G^R=-\lim_{r\to \infty} {r^3 f\frac{A'_y(r)}{A_y(r)}} .\nonumber
\end{eqnarray}

\begin{figure}[ht]
\begin{center}
\begin{minipage}[c]{0.5\textwidth}
\centering\includegraphics[width=1\textwidth]{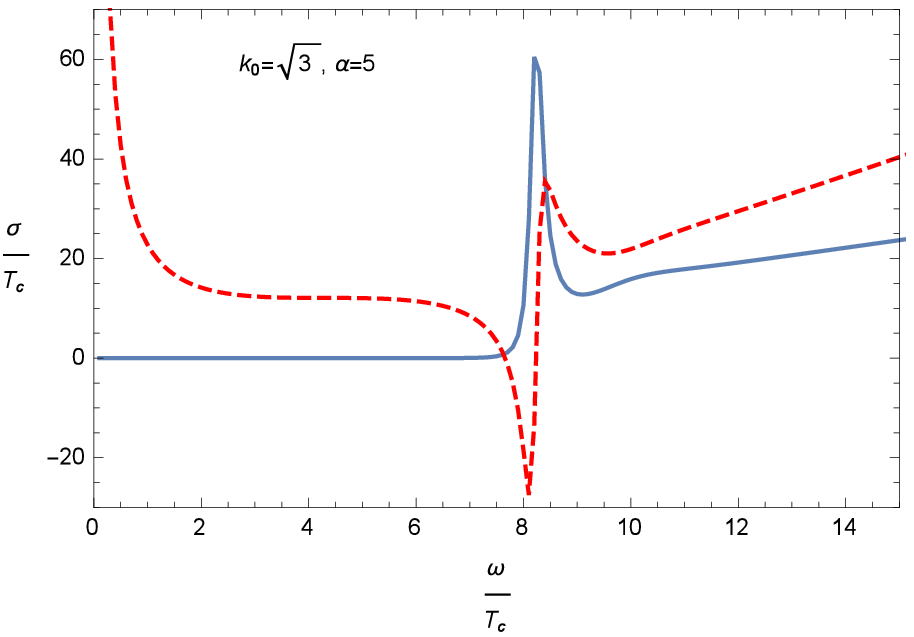}
\end{minipage}%
\begin{minipage}[c]{0.5\textwidth}
\centering\includegraphics[width=1\textwidth]{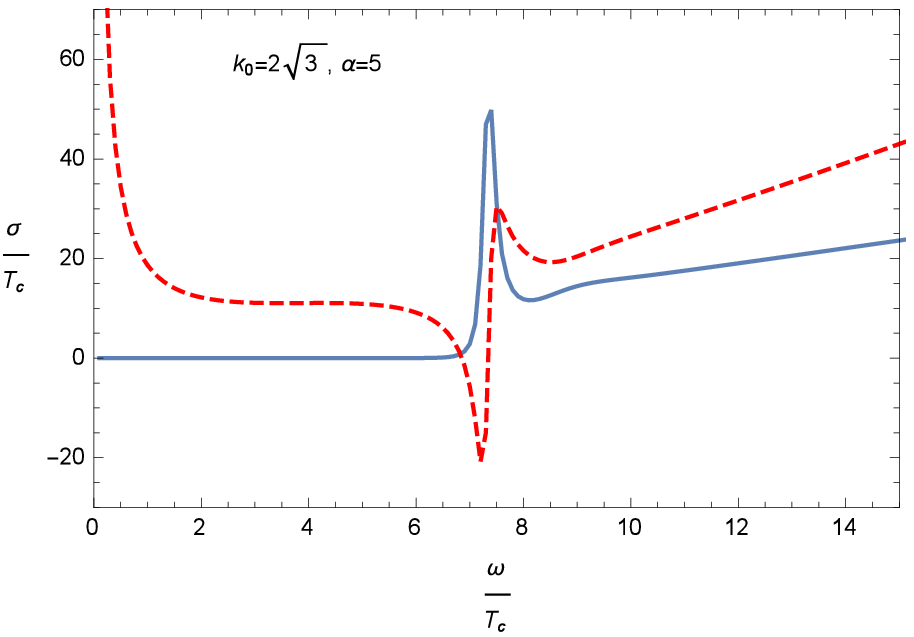}
\end{minipage}
\\
\begin{minipage}[c]{0.5\textwidth}
\centering\includegraphics[width=1\textwidth]{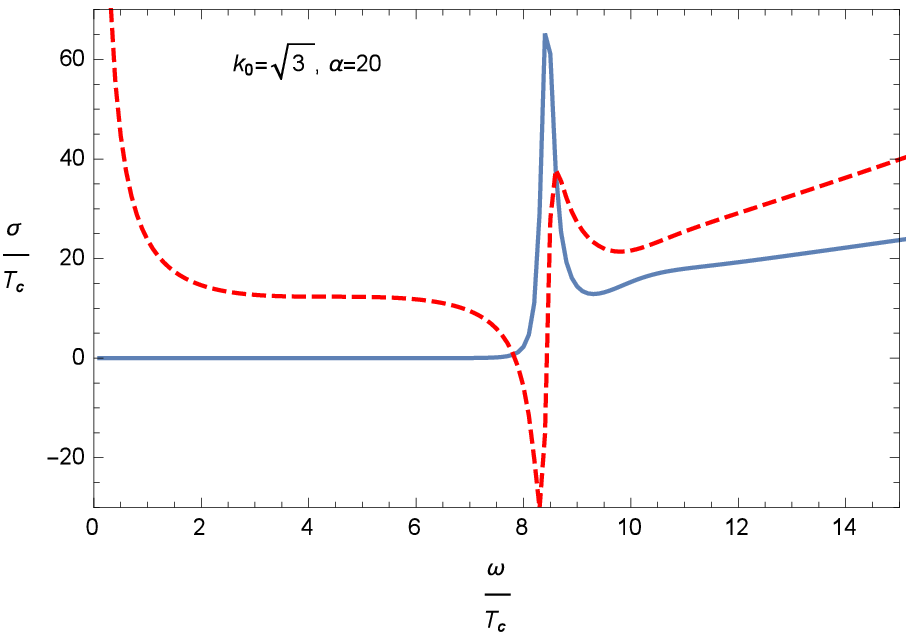}
\end{minipage}%
\begin{minipage}[c]{0.5\textwidth}
\centering\includegraphics[width=1\textwidth]{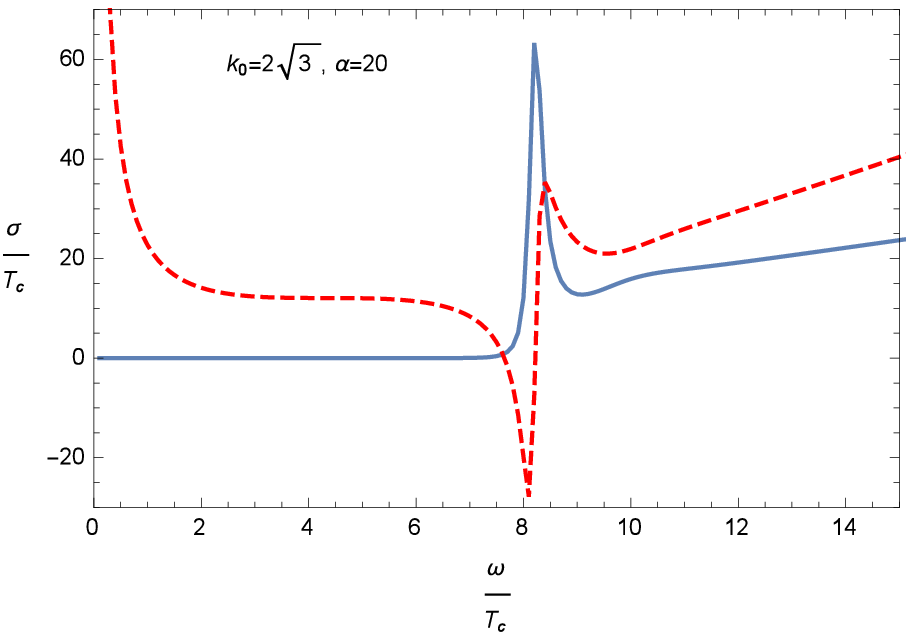}
\end{minipage}
\\
\begin{minipage}[c]{0.5\textwidth}
\centering\includegraphics[width=1\textwidth]{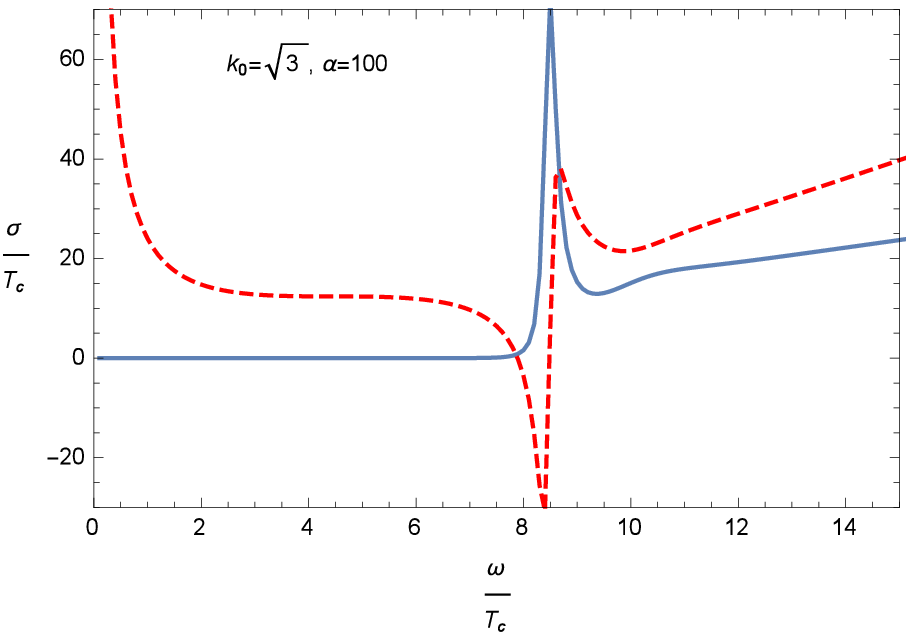}
\end{minipage}%
\begin{minipage}[c]{0.5\textwidth}
\centering\includegraphics[width=1\textwidth]{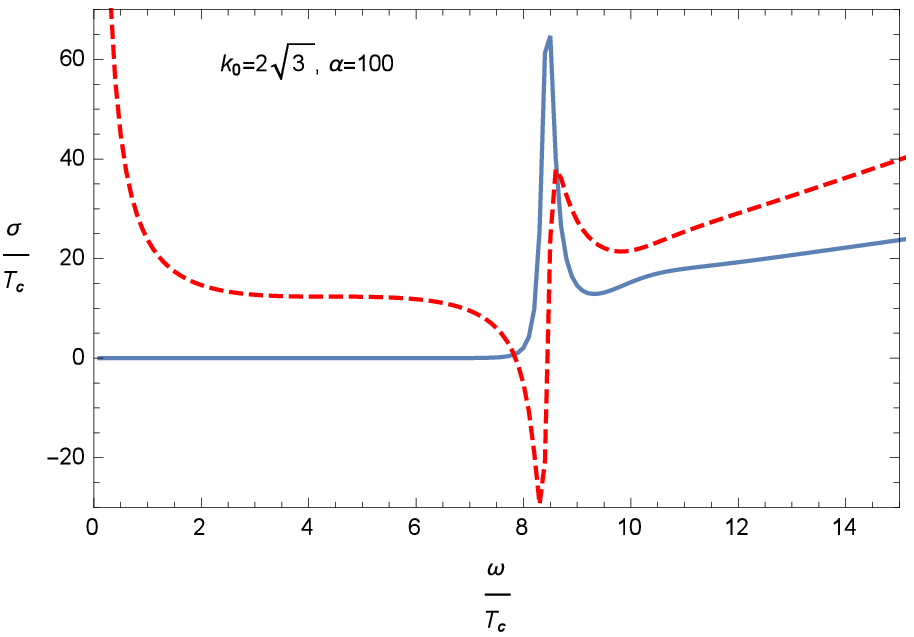}
\end{minipage}
\renewcommand{\figurename}{Fig.}
\caption{The real (blue) and imaginary (red) parts of the conductivity as a fuction of $\omega/T_c$ for $k_0=\sqrt{3}, ~2\sqrt{3}$ with $\Delta=3/2, m^2=-3/4$.}
\label{conductivity1}
\end{center}
\end{figure}

\begin{figure}[ht]
\begin{center}
\begin{minipage}[c]{0.5\textwidth}
\centering\includegraphics[width=1\textwidth]{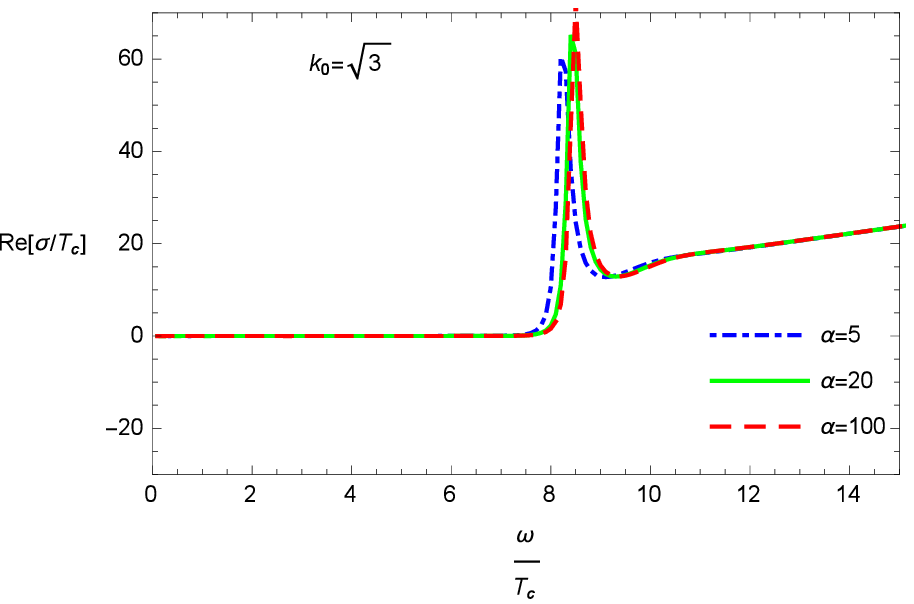}
\end{minipage}%
\begin{minipage}[c]{0.5\textwidth}
\centering\includegraphics[width=1\textwidth]{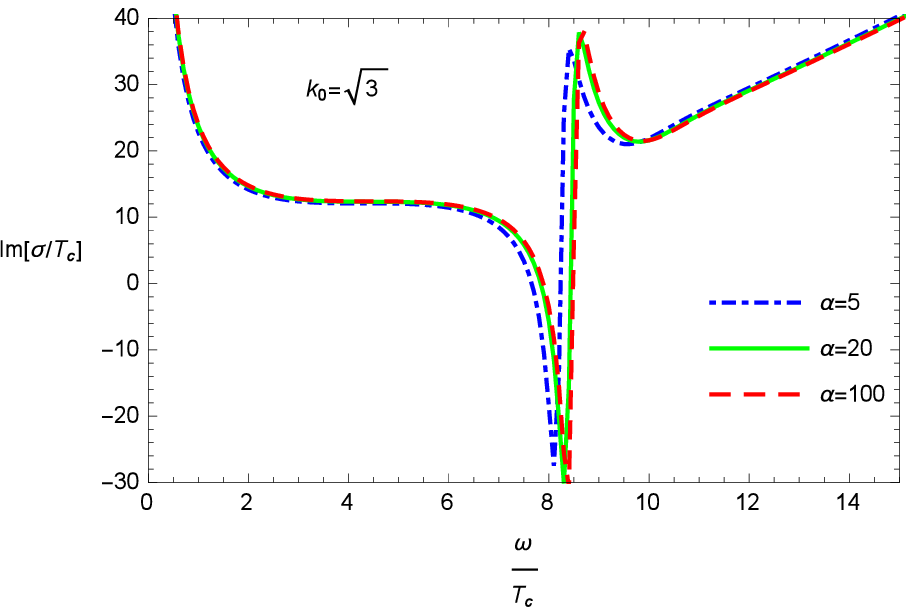}
\end{minipage}
\\
\begin{minipage}[c]{0.5\textwidth}
\centering\includegraphics[width=1\textwidth]{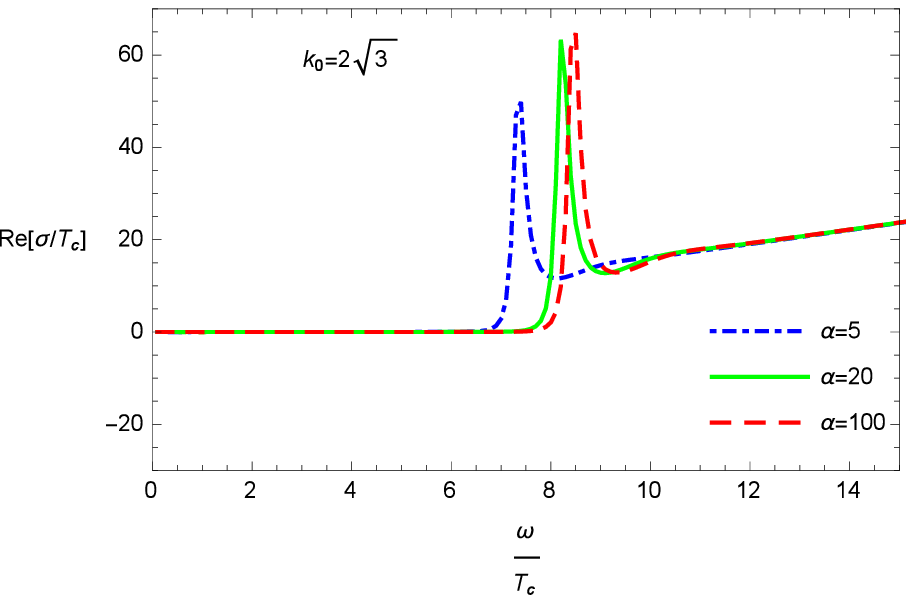}
\end{minipage}%
\begin{minipage}[c]{0.5\textwidth}
\centering\includegraphics[width=1\textwidth]{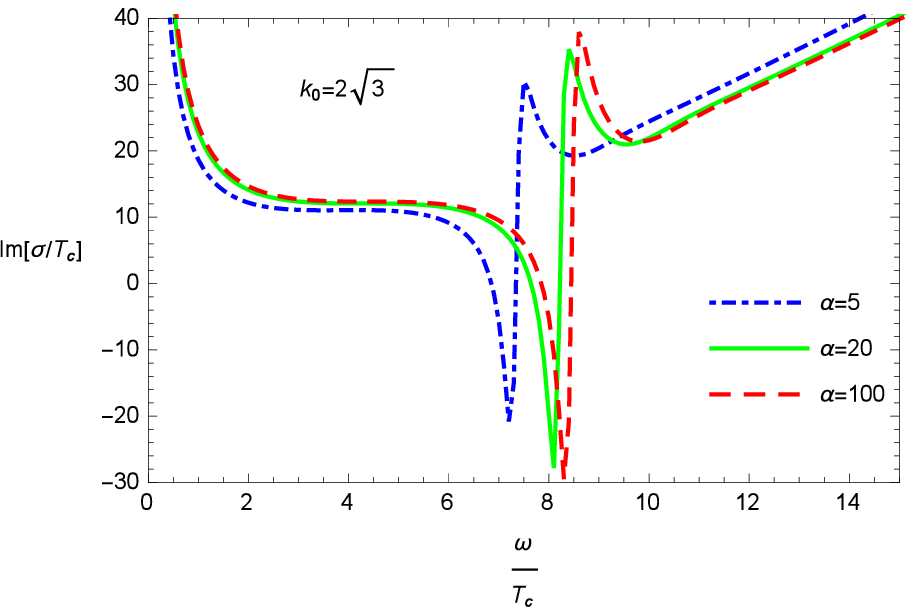}
\end{minipage}
\renewcommand{\figurename}{Fig.}
\caption{The parameter dependence of the real as well as imaginary parts of the conductivity for $k_0=\sqrt{3}, 2\sqrt{3}$ with $\Delta=3/2, m^2=-3/4$. }
\label{conductivity2}
\end{center}
\end{figure}

By solving Eq.(\ref{eqAy}) numerically, one obtains the conductivity as a function of the ratio between the frequency to the temperature for the different parameters which is presented in Fig.\ref{conductivity1}.
In the plots, the solid blue curves represent the real part, and the dashed red curves represent the imaginary part of the conductivity respectively.
The calculations are carried out with different $k_0$, $\alpha$, and $T/T_c$ for given $m$ as well as $\Delta$.
To analyze its dependence on $\alpha$, we reorganize the results and show them in Fig.\ref{conductivity2}.

%Following Refs.~\cite{horowitz2008holographic,lu2016holographic}, we define $\omega_g$ as the frequency of the first pole on the right hand side of $\omega=0$ which minimizes the imaginary part of the conductivity when $\Delta>\Delta_{BF}$ and $m^2>m_{BF}^2$, and calculate the ratio of $\omega_g$ to the critical temperature $\omega_g/T_c$.
Following Refs.~\cite{horowitz2008holographic,lu2016holographic}, we define $\omega_g$ as the frequency which minimizes the imaginary part of the conductivity when $\Delta>\Delta_{BF}$ and $m^2>m_{BF}^2$, and calculate the ratio of $\omega_g$ to the critical temperature $\omega_g/T_c$.
From these figures, one observes that $\omega_g/T_c$ does not remain unchanged for different parameters, but rather varies as functions of $k_0$ and $\alpha$ in the present model.
Nevertheless, it turns out to be larger than the BCS value $3.5$.

As we can see from Fig.\ref{conductivity1} and Fig.\ref{conductivity2}, the conductivity rises sharply at $\omega_g/T_c$.
The value of $\omega_g/T_c$ varies with different $k_0$ and $\alpha$.
For $\alpha=5$, the gap $\omega_g/T_c$ in the case of $k_0=\sqrt{3}$ is larger than that in the case of $k_0=2\sqrt{3}$.
Moreover, the peak of the real part of the conductivity for $k_0=\sqrt{3}$ is higher than that for $k_0=2\sqrt{3}$.
With the increase of $\alpha$, the peak of the real part of the conductivity and $\omega_g/T_c$ also increase.
For $\alpha=20$, the peak of the real part of the conductivity with $k_0=2\sqrt{3}$ starts to approach that with $k_0=\sqrt{3}$.
As $\alpha$ further increases, the gap $\omega_g/T_c$ for $k_0=\sqrt{3}$ and $k_0=2\sqrt{3}$ becomes numerically identical.

One finds, from Fig.\ref{conductivity2}, that the gap $\omega_g/T_c$ increases distinctly with increasing $\alpha$ for $k_0=2\sqrt{3}$, which implies that a smaller value of $\alpha$ is favorable to the superconductor phase.
Equivalently, owing to the existence of the scalar charge $\tilde{q}$, the phase transition is easier to happen in the hairy planar black holes than in Schwarzschild-AdS which corresponds to the limit $\alpha\to\infty$.
Besides, the value of $k_0$ also has an impact on the gap, and $\omega_g/T_c$ becomes smaller with increasing $k_0$ for given values of $\alpha$.
%The conductivity curve is observed to stabilize quickly for parameters $\alpha=5,~20,~100$, toward the right of the pole presented by the spike at $\omega_g$.
The conductivity curve is observed to stabilize quickly for parameters $\alpha=5,~20,~100$, toward the right of the spike at $\omega_g$.
In particular, as $\alpha$ increases, the entire curve of conductivity tends to converge to a given form.
Therefore, one concludes that, for moderate temperature, the gap $\omega_g/T_c$ is sensitive to the value of $k_0$ only for smaller $\alpha$.
We also find that the peaks of the real part of the conductivity increase when $\alpha$ increases moderately.
But as $\alpha$ increases, the resultant conductivity becomes insensitive to other parameters.

\begin{figure}[ht]
\begin{center}
\begin{minipage}[c]{0.5\textwidth}
\centering\includegraphics[width=1\textwidth]{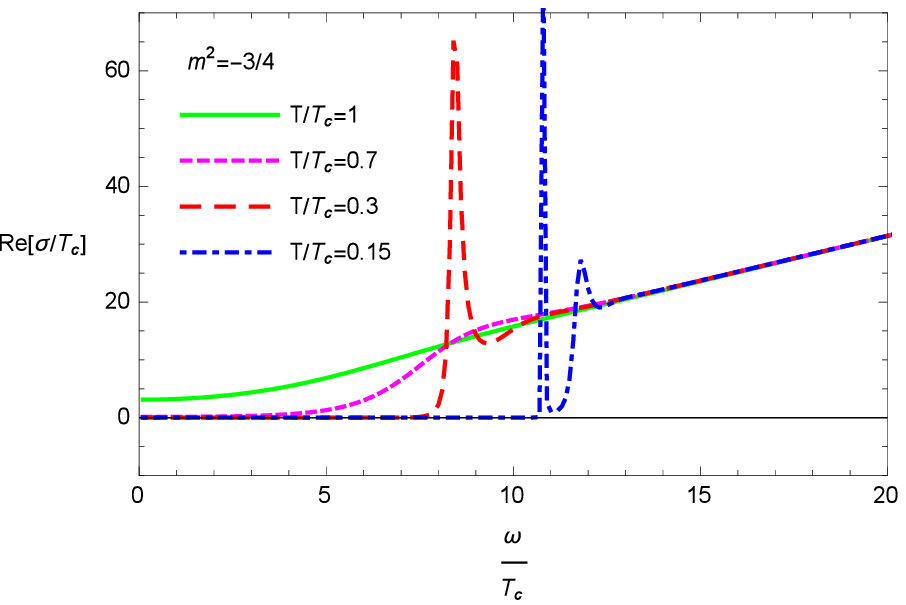}
\end{minipage}%
\begin{minipage}[c]{0.5\textwidth}
\centering\includegraphics[width=1\textwidth]{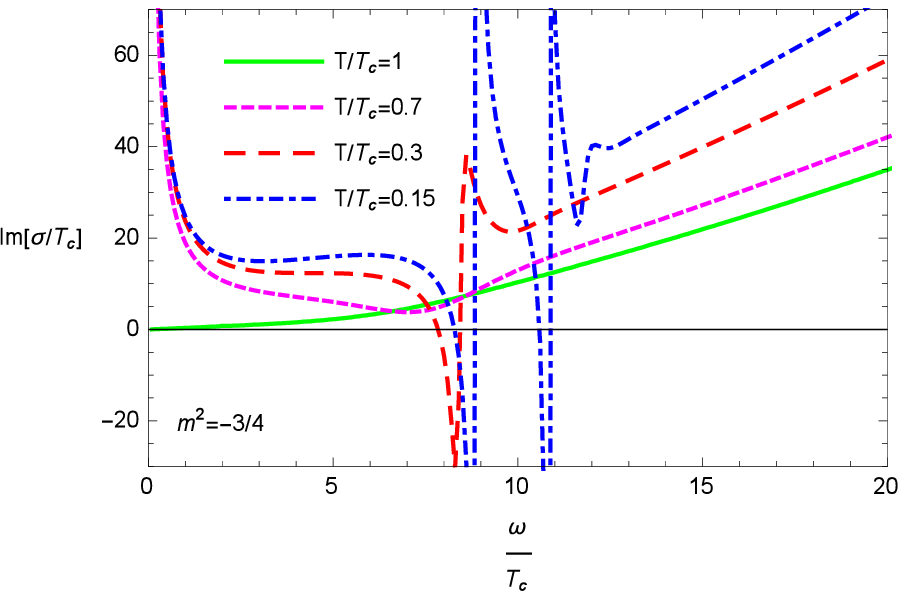}
\end{minipage}
\\
\begin{minipage}[c]{0.5\textwidth}
\centering\includegraphics[width=1\textwidth]{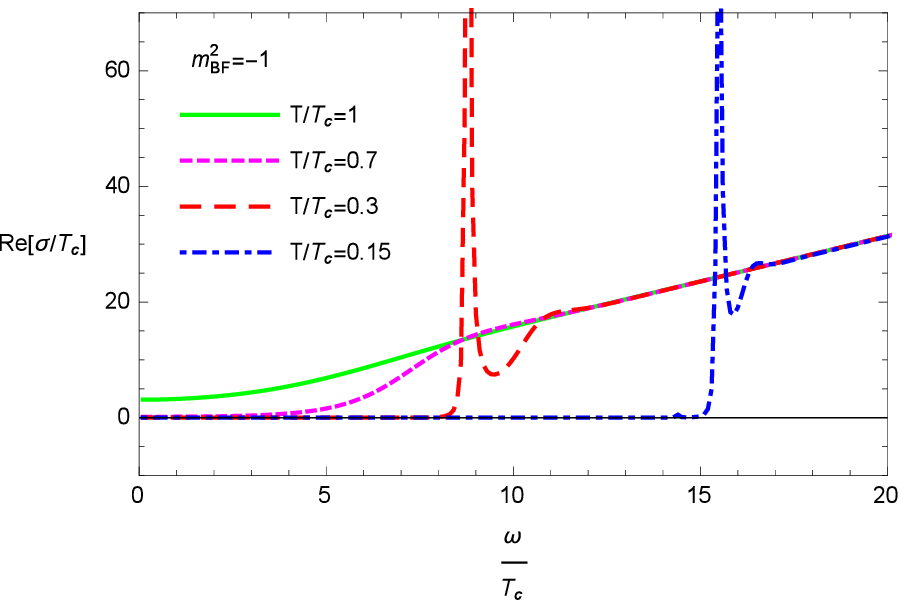}
\end{minipage}%
\begin{minipage}[c]{0.5\textwidth}
\centering\includegraphics[width=1\textwidth]{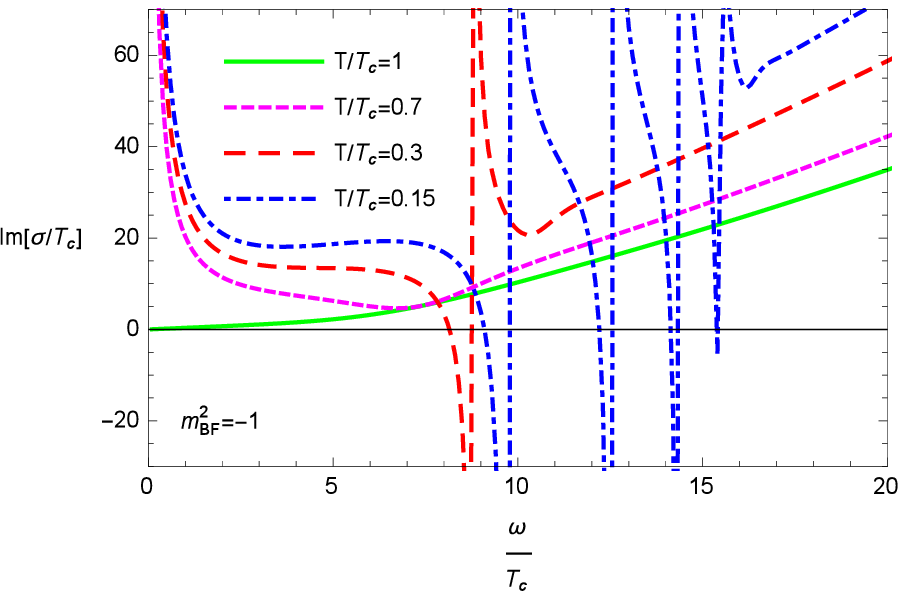}
\end{minipage}
\renewcommand{\figurename}{Fig.}
\caption{The temperature dependence of the real as well as imaginary parts of the conductivity for $m^2=-3/4, m_{BF}^2=-1$ with $k_0=\sqrt{3}$ and $\alpha=20$.}
\label{conductivity3}
\end{center}
\end{figure}

Last but not least, we study the temperature dependence of the conductivity in Fig.\ref{conductivity3}. 
The calculations were carried out by taking $k_0=\sqrt{3}$ and $\alpha=20$ for diferent values of the mass of the vector field $m^2=-3/4$ and $m^2=m_{BF}^2=-1$. 
The real part of the conductivity is found to be finite shortly after the superconducting phase is formed, namely, $T/T_{c} \lesssim 1$. 
This indicates a finite contribution to the normal component of the conductivity, and therefore, the system can be interpreted in terms of the two-fluid model~\cite{adscft-superconductor-07}. 
On the other hand, as the temperature decreases, especially when $T/T_{c} < 0.3$, the real part of the conductivity mostly vanishes.
In other words, the soft gap in the low-frequency region gradually transforms into a superconducting gap as the temperature becomes sufficiently small regarding the frequencies between $\omega=0$ and $\omega=\omega_g$.
This is quantitatively in agreement to the findings in Ref.\cite{adscft-superconductor-07}.
Moreover, from Fig.\ref{conductivity3} we find that additional poles show up as the temperature decreases to a certain extent, for instance, when $T/T_{c}=0.15$ and $m^2=-3/4$. 
This indicates that, when the mass of the vector field is close to the BF bound, the multiple poles can be observed at a sufficiently low temperature. 
As expected, the same feature is also readily observed when the mass of the vector field is identical to the BF bound, $m^2=-1$.
These results are in agreement with the standard s-wave superconductor for the Schwarzschild black hole~\cite{horowitz2008holographic}.

%%%%%%%%%%%%%%%%%%%%%
\section{V Concluding remarks}
%%%%%%%%%%%%%%%%%%%%%

To summarize, we have investigated the properties of the p-wave superconductor in the background  metric of hairy black holes.
These black hole solutions are derived from the Einstein gravity minimally coupled to a scalar field with a generic potential.
The solutions are characterized by two independent parameters which might restore the AdS vacuum in particular circumstances.
By using the hairy black hole solution as background, we have investigated the properties of the conductor/superconductor phase transition of the dual system.
We have studied various physical quantities concerning the transition regarding the parameters of the background metric.

It is shown that the presence of a scalar hair for the background black hole metric, in terms of model parameters, has a significant impact on the properties of the holographic superconductor.
The effects of the parameters $\alpha$ and $k_0$ on the critical temperature are studied.
For a moderate value of temperature and a given $k_0$, the critical temperature decreases with the increase of $\alpha$ and approaches a universal constant for $\alpha\to \infty$.
These features can be understood in terms of the properties of the effective mass of the vector field.
It is found that the critical temperature depends on the effective mass.
The model parameter dependence of the condensation, in turn, inherits from the corresponding dependence of the effective mass.
We also find that the conductor/superconductor phase transition is more difficult to take place for large $\alpha$ and small $k_0$.
In comparison with the condensation in the Schwarzschild-AdS black hole background, which can be restored in the present model by taking $\alpha\to\infty$,  the superconductor transition is easier to take place owing to the presence of the scalar hair $\Phi$.

The calculations of the conductivity in five-dimensional black holes for different $k_0$ are also carried out.
The value of the gap $\omega_g/T_c$ is found to vary as a function of the parameters $\alpha$ and $k_0$.
Nonetheless, the obtained value is still more significant than that achieved in the BCS theory, which suggests that the dual system indeed involves strong interactions.
Also, by studying the temperature dependence of the conductivity, the gap is found to be a soft gap as the contribution to the normal component of the conductivity is finite.
Another feature is the emergence of additional poles as the mass of the vector field approach to the Breitenlohner-Freedman bound and the temperature becomes sufficiently small.
This indicates that further stable quasiparticle states might have been formed.
Lastly, by evaluating the free energies of the superconducting and normal phases, the former is found to be thermodynamically favorable.

%%%%%%%%%%%%%%%%%%%%%%
\section*{Acknowledgements}
%%%%%%%%%%%%%%%%%%%%%%%
We are thankful for valuable discussions with Yunqi Liu, Xiaomei Kuang, and Chenyong Zhang.
We gratefully acknowledge the financial support from Brazilian funding agencies Funda\c{c}\~ao de Amparo \`a Pesquisa do Estado de S\~ao Paulo (FAPESP),
Conselho Nacional de Desenvolvimento Cient\'{\i}fico e Tecnol\'ogico (CNPq), Coordena\c{c}\~ao de Aperfei\c{c}oamento de Pessoal de N\'ivel Superior (CAPES),
and National Natural Science Foundation of China (NNSFC) under contract Nos.11690034, 11775076, 11573022 and 11375279,
as well as Hunan Provincial Natural Science Foundation of China under Grant No. 2016JJ1012.

\bibliographystyle{h-physrev}
\bibliography{reference_dan,references_qian}

\begin{thebibliography}{10}

\bibitem{hooft1993dimensional}
G.~'t~Hooft,
\newblock Conf. Proc. {\bf C930308}, 284 (1993), arXiv:9310026.

\bibitem{susskind1995world}
L.~Susskind,
\newblock J. Math. Phys. {\bf 36}, 6377 (1995), arXiv:9409089.

\bibitem{maldacena1999large}
J.~M. Maldacena,
\newblock AIP Conf. Proc. {\bf 484}, 51 (1999).

\bibitem{adscft-rhic-01}
J.~Casalderrey-Solana, H.~Liu, D.~Mateos, K.~Rajagopal, and U.~A. Wiedemann,
\newblock (2011), arXiv:1101.0618.

\bibitem{adscft-condensed-review-02}
C.~P. Herzog,
\newblock J. Phys. {\bf A42}, 343001 (2009), arXiv:0904.1975.

\bibitem{gubser2008breaking}
S.~S. Gubser,
\newblock Phys. Rev. D {\bf 78}, 065034 (2008).

\bibitem{hartnoll2008building}
S.~A. Hartnoll, C.~P. Herzog, and G.~T. Horowitz,
\newblock Phys. Rev. Lett. {\bf 101}, 031601 (2008), arXiv:0803.3295.

\bibitem{chen2010towards}
J.-W. Chen, Y.-J. Kao, D.~Maity, W.-Y. Wen, and C.-P. Yeh,
\newblock Phys. Rev. D {\bf 81}, 106008 (2010).

\bibitem{benini2010gauge}
F.~Benini, C.~P. Herzog, R.~Rahman, and A.~Yarom,
\newblock JHEP {\bf 2010}, 137 (2010).

\bibitem{mackenzie2000p}
A.~Mackenzie and Y.~Maeno,
\newblock Physica B: Condensed Matter {\bf 280}, 148 (2000).

\bibitem{cai2013holographic}
R.-G. Cai, S.~He, L.~Li, and L.-F. Li,
\newblock JHEP {\bf 2013}, 36 (2013).

\bibitem{horowitz2011introduction}
G.~T. Horowitz,
\newblock Lect. Notes Phys. {\bf 828}, 313 (2011), arXiv:1002.1722.

\bibitem{herzog2009lectures}
C.~P. Herzog,
\newblock J. Phys. {\bf A42}, 343001 (2009), arXiv:0904.1975.

\bibitem{Lin:2016swr}
K.~Lin, A.~B. Pavan, Q.~Pan, and E.~Abdalla,
\newblock Braz. J. Phys. {\bf 46}, 767 (2016).

\bibitem{Pan:2015lit}
Q.~Pan and S.-J. Zhang,
\newblock Eur. Phys. J. {\bf C76}, 126 (2016), arXiv:1510.09199.

\bibitem{Wu:2014bba}
Y.-B. Wu {\em et~al.},
\newblock Phys. Rev. {\bf D90}, 126006 (2014), arXiv:1410.5243.

\bibitem{Lan:2017qxm}
S.~Lan, W.~Liu, and Y.~Tian,
\newblock Phys. Rev. {\bf D95}, 066013 (2017), arXiv:1701.02921.

\bibitem{Pan:2011ah}
Q.~Pan, J.~Jing, and B.~Wang,
\newblock JHEP {\bf 11}, 088 (2011), arXiv:1105.6153.

\bibitem{Cai:2010cv}
R.-G. Cai, Z.-Y. Nie, and H.-Q. Zhang,
\newblock Phys. Rev. {\bf D82}, 066007 (2010), arXiv:1007.3321.

\bibitem{Gangopadhyay:2012np}
S.~Gangopadhyay and D.~Roychowdhury,
\newblock JHEP {\bf 05}, 156 (2012), arXiv:1204.0673.

\bibitem{wen2018maxwell}
D.~Wen, H.~Yu, Q.~Pan, K.~Lin, and W.-L. Qian,
\newblock Nucl. Phys. {\bf B930}, 255 (2018), arXiv:1803.06942.

\bibitem{lu2016holographic}
J.-W. Lu {\em et~al.},
\newblock Nucl. Phys. B {\bf 903}, 360 (2016).

\bibitem{Nishioka:2009zj}
T.~Nishioka, S.~Ryu, and T.~Takayanagi,
\newblock JHEP {\bf 03}, 131 (2010), arXiv:0911.0962.

\bibitem{Surya:2001vj}
S.~Surya, K.~Schleich, and D.~M. Witt,
\newblock Phys. Rev. Lett. {\bf 86}, 5231 (2001), arXiv:0101134.

\bibitem{Mansoori:2016zbp}
S.~A. Hosseini~Mansoori, B.~Mirza, A.~Mokhtari, F.~L. Dezaki, and
  Z.~Sherkatghanad,
\newblock JHEP {\bf 07}, 111 (2016), arXiv:1602.07245.

\bibitem{fan2015static}
Z.-Y. Fan and H.~L{\"u},
\newblock Phys. Rev. D {\bf 92}, 064008 (2015).

\bibitem{fan2016exact}
Z.-Y. Fan and B.~Chen,
\newblock Phys. Rev. D {\bf 93}, 084013 (2016).

\bibitem{cai2015introduction}
R.-G. Cai, L.~Li, L.-F. Li, and R.-Q. Yang,
\newblock Sci. China Phys. Mech. Astron. {\bf 58}, 060401 (2015),
  arXiv:1502.00437.

\bibitem{hartnoll2008holographic}
S.~A. Hartnoll, C.~P. Herzog, and G.~T. Horowitz,
\newblock JHEP {\bf 2008}, 015 (2008).

\bibitem{adscft-superfluid-02}
C.~P. Herzog, P.~K. Kovtun, and D.~T. Son,
\newblock Phys. Rev. {\bf D79}, 066002 (2009), arXiv: 0809.4870.

\bibitem{adscft-superfluid-04}
D.~Arean, P.~Basu, and C.~Krishnan,
\newblock JHEP {\bf 10}, 006 (2010), arXiv: 1006.5165.

\bibitem{adscft-superfluid-06}
S.~Liu, Q.~Pan, and J.~Jing,
\newblock Phys. Lett. {\bf B765}, 91 (2017), 1610.02549.

\bibitem{gubser1998gauge}
S.~S. Gubser, I.~R. Klebanov, and A.~M. Polyakov,
\newblock Phys. Lett. B {\bf 428}, 105 (1998).

\bibitem{gubser2008gravity}
S.~S. Gubser and S.~S. Pufu,
\newblock JHEP {\bf 2008}, 033 (2008).

\bibitem{Chaturvedi:2015hra}
P.~Chaturvedi and G.~Sengupta,
\newblock JHEP {\bf 04}, 001 (2015), arXiv:1501.06998.

\bibitem{horowitz2008holographic}
G.~T. Horowitz and M.~M. Roberts,
\newblock Phys. Rev. D {\bf 78}, 126008 (2008).

\bibitem{adscft-superconductor-07}
J.-P. Wu and P.~Liu,
\newblock Phys. Lett. {\bf B774}, 527 (2017), arXiv:1710.07971.

\end{thebibliography}

\end{document}